\documentclass[onecolumn]{aa} 
\usepackage{graphicx,amsmath,deluxetable}
\usepackage{txfonts}
\usepackage{natbib}
\bibpunct{(}{)}{;}{a}{}{,} 
\begin{document}
   \authorrunning{G.~W.~Fuchs et al.}
   \titlerunning{Hydrogenation reactions in interstellar CO ice analogues}
\title{Hydrogenation reactions in interstellar CO ice analogues}

\subtitle{a combined experimental/theoretical approach}

\author{G.~W.~Fuchs, H.~M.~Cuppen, S.~Ioppolo, C.~Romanzin, S.~E.~Bisschop, S.~Andersson, E.~F.~van Dishoeck, and H.~Linnartz}

   \offprints{H.~Linnartz; \email{linnartz@strw.leidenuniv.nl}}

   \institute{Raymond \& Beverly Sackler Laboratory for Astrophysics,
Leiden Observatory, Leiden University, PO Box 9513, 2300 RA Leiden,
              The Netherlands}

    \date{Received ; accepted }


  \abstract
   {Hydrogenation reactions of CO in inter- and circumstellar ices are regarded as an important starting point in the formation of more complex species. Previous laboratory measurements by two groups on the hydrogenation of CO ices resulted in controversial results on the formation rate of methanol (2002, ApJ, 577, 265 and 2002, ApJL, 571, L173). }
   {Our aim is to resolve this controversy by an independent investigation of the reaction scheme for a range of H-atom fluxes and different ice  temperatures and thicknesses. In order to fully understand the laboratory data, the results are interpreted theoretically by means of continuous-time, random-walk Monte Carlo simulations. }
   {Reaction rates are determined by using a state-of-the-art ultra high vacuum experimental setup to bombard an interstellar CO
ice analog with room temperature H atoms. The reaction of CO + H into H$_2$CO and subsequently CH$_3$OH is monitored by a Fourier transform
infrared spectrometer in a reflection absorption mode. In addition, after each completed measurement a temperature programmed
desorption experiment is performed to identify the produced species according to their mass spectra and to determine their
abundance. Different H-atom fluxes, morphologies, and ice thicknesses are tested. The experimental results are
interpreted using Monte Carlo simulations. This technique takes into account the layered structure of CO ice. }
{ The formation of both formaldehyde and methanol via CO hydrogenation
  is confirmed at low temperature ($T = 12-20$ K). We confirm, as
  proposed by Hidaka et al. (2004, ApJ, 614, 1124), that the
  discrepancy between the two Japanese studies is mainly due to a
  difference in the applied hydrogen atom flux. The production rate of
  formaldehyde is found to decrease and the penetration column to
  increase with temperature. {Temperature-dependent }reaction barriers and 
  diffusion rates are inferred using a Monte Carlo physical chemical model. The model is extended to interstellar
  conditions to compare with observational H$_2$CO/CH$_3$OH data.}
   {}


   \maketitle

\section{Introduction}
In recent years an increasing number of experimental and theoretical studies have been focussing on the characterisation of solid state astrochemical processes. These studies were triggered by the recognition that many of the simple and more complex molecules in the interstellar medium are most likely formed on the surfaces of dust grains. Astronomical observations along with detailed laboratory studies and recent progress in UHV surface techniques have made possible an experimental verification of the initial surface reaction schemes introduced by Tielens, Hagen and Charnley \citep{Tielens:1982, Tielens:1997}. Very recently the formation of water was shown in hydrogenation schemes starting from solid molecular oxygen \citep{Miyauchi:2008,Ioppolo:2008} and that of ethanol from acetaldehyde  \citep{Bisschop:2007I}. The first solid state astrochemical laboratory studies focused on the formation of  formaldehyde and methanol by H-atom bombardment of CO ice.  
Methanol is abundantly observed in interstellar ices and is considered to be a resource for the formation of more complex molecules via surface reactions and after evaporation in the gas phase \citep{Charnley:1992}. 
The hydrogenation scheme for the solid state formation of methanol was proposed as
\begin{equation}
\rm
CO \xrightarrow{H} HCO  \xrightarrow{H} H_2CO  \xrightarrow{H} H_3CO  \xrightarrow{H} CH_3OH
\end{equation}

The past laboratory studies of H-atom bombardment of CO ice have been performed independently by two groups \citep{Hiraoka:2002, Watanabe:2002}. 
 \cite{Hiraoka:2002} observed only formaldehyde formation, whereas \cite{Watanabe:2002} also found efficient methanol production. In a series of papers these conflicting results have been discussed \citep{Hiraoka:2002, Watanabe:2003,Watanabe:2004} and the existing discrepancy has been proposed as a consequence of different experimental conditions, most noticeable the adopted H-atom flux \citep{Hidaka:2004}. 
Understanding the solid state formation route to methanol became even more pressing with the recent experimental finding that the gas-phase formation route via ion-neutral reactions is  less efficient than thought before and cannot explain the observed interstellar abundances \citep{Geppert:2005, Garrod:2006}. 

Recently, also deuteration experiments were performed on CO ice which confirmed the formation of both fully deuterated formaldehyde and methanol, but with substantially lower reaction rates \citep{Nagaoka:2005, Watanabe:2006}. It was suggested that in the presence of both hydrogen and deuterium first the normal methanol  forms which then gradually converts to the deuterated species by exchange reactions.

 The present paper strongly supports the flux argument given by \cite{Hidaka:2004}.
It furthermore presents a systematic study of the physical dependencies involved in the CO-ice hydrogenation with the aim to put previous work in a context that allows an extension of solid state astrochemical processes to more complex species. Special emphasis is put on the flux and temperature dependence of the formation rate. An analysis of the spectral changes of CO ice during hydrogenation is included to give insight in the structure of the reactive layer. Furthermore, Monte Carlo simulations are presented that allow to interpret the experimental results in more detail and to vary parameters that are hard to study independently by  experiment. We conclude with a simulation of  H$_2$CO/CH$_3$OH formation under interstellar conditions, in particular for low H-atom fluxes. The outcome is compared with astronomical observations.
\section{Experimental procedure}
The experiments are performed under UHV conditions. The room temperature base pressure of the vacuum system is better than 3 $\times$ 10$^{-10}$ mbar. Figure~\ref{setup} shows a schematic representation of the setup. (See \cite{Ioppolo:2008} for additional information) Amorphous CO ices ranging from a few to several monolayers are grown on a gold coated copper substrate that is located in the centre of the main chamber and mounted on the tip of a cold finger of a 10 K He cryostat. The temperature of the ice is controlled between 12.0 K and 300 K with 0.5 K relative precision between experiments. The absolute accuracy is better than 2 K. During deposition the layer thickness is monitored by simultaneous recording of reflection absorption infrared (RAIR) spectra. In order to exclude the effect of potential pollutions, ices are grown using CO, $^{13}$CO or C$^{18}$O isotopologues.  

The ice layers are exposed to a hydrogen atom beam. The atoms are produced by a well characterised commercial thermal cracking source \citep{Tschersich:1998,Tschersich:2000} that provides H-atom fluxes on the sample surface between 10$^{12}$ and 10$^{14}$ atoms cm$^{-2}$s$^{-1}$. For comparison, the Hiraoka group used fluxes below 10$^{13}$ atoms cm$^{-2}$s$^{-1}$ and the Watanabe group worked in the 10$^{14}$-10$^{15}$ atoms cm$^{-2}$s$^{-1}$ regime. The hot ($\sim$ 2000 K) hydrogen atoms are cooled down to room temperature via surface collisions in a nose-like shape quartz pipe between the atomic source and the ice sample. In this way hot hydrogen atoms cannot affect the ice directly. H-atom recombination in this connecting pipe results in a lower final flux.  {Details about the flux determination are given in Appendix~A. The absolute fluxes are estimated to be within a factor of two, the relative fluxes within 50~\%.}

{The relatively high temperature of the incident atoms of 300 K} does not affect the process; previous experiments with colder H atoms did not show any substantial temperature dependence because the atoms are immediately thermalized on the surface \citep{Watanabe:2002}. It is argued that the surface is covered with a thin layer of hydrogen molecules under these conditions. These molecules are either formed on the surface or originate from the partially dissociated beam. Since the incoming atoms have to penetrate this cold H$_2$ layer, they are thermally adjusted to the surface temperature once they come in contact with the CO molecules. 

Information about the reaction products is obtained using two complementary techniques.  During the H-atom bombardment reactants and products are monitored by recording RAIR spectra. The RAIR spectra are recorded using a Fourier transform infrared spectrometer with 1 and 4 cm$^{-1}$ resolution and covering the spectral region in which CO (2143 (s) cm$^{-1}$), formaldehyde (1732~(s), 1479 and 2812~(m), and 1246, 1175, 2991, 2880, and 2812~(mw)~cm$^{-1}$) and methanol (1035 (s) and 1125 (w) cm$^{-1}$) exhibit strong (s), medium (m) or weak (w) absorptions. The intensity of spectral features is directly related to the density in the ice. The products are monitored mass spectrometrically using temperature programmed desorption (TPD) once a hydrogenation experiment is completed.

\section{Experimental results}
\subsection{A sample experiment}

To illustrate the experimental method we start by discussing a sample experiment in which a CO ice of 8$\times$10$^{15}$ molecules cm$^{-2}$ is bombarded with H atoms with a flux of 5$\times$10$^{13}$ cm$^{-2}$ s$^{-1}$ for three hours at a surface temperature of 12.0 K. This corresponds to a fluence of 5.4$\times$10$^{17}$ cm$^{-2}$.
Figure~\ref{IR_hiflux} shows the RAIR difference spectrum {($\Delta Abs$)} after these three hours of exposure (after~$-$~before). Indicated are the CO, the H$_2$CO and the CH$_3$OH spectral signatures with respect to the spectrum recorded before the H-atom bombardment started. The CO appears as a negative band indicating its use-up and the other bands are positive, indicating the formation of H$_2$CO and CH$_3$OH. Neither the intermediate species, HCO and H$_3$CO, nor more complex species are observed. 

\begin{figure*}[htp]
\begin{center}
\includegraphics[width=0.9\textwidth]{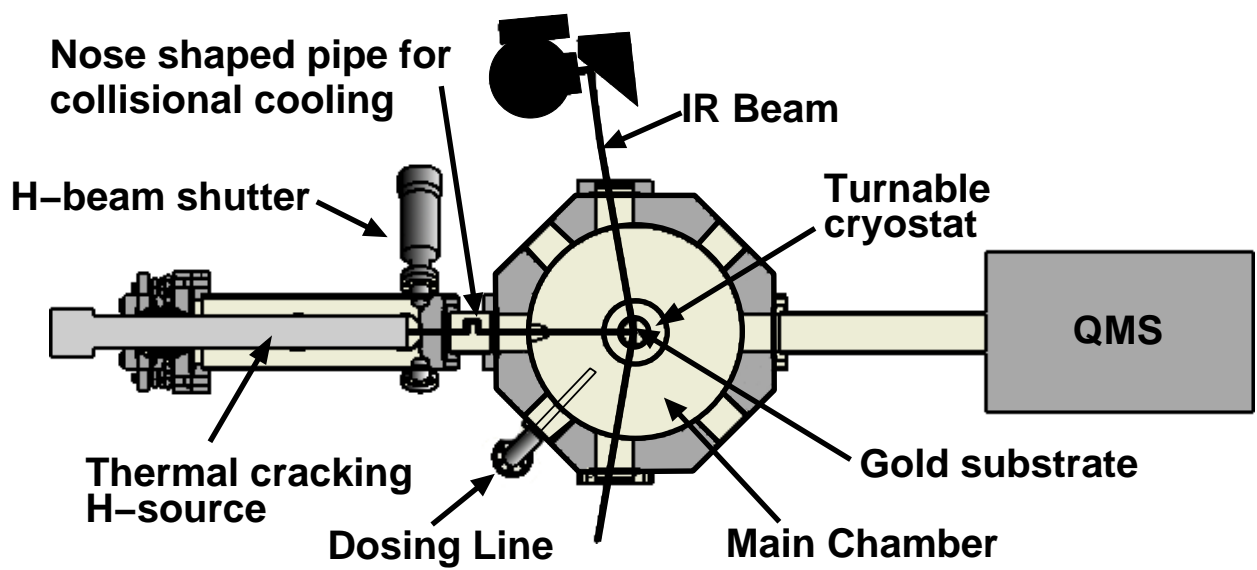}
\end{center}
\caption{Schematic representation of the experimental setup. {CO ice is deposited through the dosing line and the products are monitored by means of infrared spectroscopy and quadrupole mass spectroscopy (QMS).} }
\label{setup}
\end{figure*}

The
column density $N_{X}$ (molecules cm$^{-2}$) of 
species $X$ in the ice is calculated using
\begin{equation}
N_{X}=\frac{\int A(\nu) \textrm{d} \nu}{S_{X}} \label{N_X}
\end{equation}
where $A(\nu)$ is the wavelength dependent absorbance. Since literature values of transmission
band strengths cannot be used in reflection measurements, an
apparent absorption band strength, $S_{X}$ of species $X$ is
calculated from a calibration experiment in which an ice layer of
species $X$  desorbs at constant temperature until the
sub-monolayer regime. This is illustrated in Fig.~\ref{Calibration} that shows the decrease in integrated absorbance of CO and CH$_3$OH during such an experiment. The arrows in the graph indicate the deviation onset from constant desorption which marks the transition point from multi- to sub-monolayer regime. The thus obtained apparent absorption band strengths of CO and CH$_3$OH (1035 cm$^{-1}$) are setup specific.
The corresponding uncertainty in the band strengths remains within 50~\%. The ratio between $S_{\rm CO}$ and $S_{\rm CH_3OH}$ in our reflection experiment is similar to the transmittance ratio, 0.85.  The value for $S_{\rm H_2CO}$ is obtained by assuming mass balance 
\begin{equation}
N_{\rm CO}(t) + N_{\rm CH_3OH}(t) = - \frac{\int A(\nu) \textrm{d} \nu}{S_{\rm H_2CO}} 
\end{equation}
for a set of different experiments. In addition, the results discussed in the present paper are all in a regime where the proportionality relation \citep{Teolis:2007} still holds ($<$ 3$\times 10^{16}$ molecules cm$^{-2}$).

The CO band shape can change when more molecules other than CO are formed. Figure~\ref{COchange} shows the 2143 cm$^{-1}$ IR peak before and after the H-atom exposure. A clear decrease of the peak height can be observed due to the use-up of CO during the experiment, as is expected. However, an additional peak appears at 2135 cm$^{-1}$ (see inset Fig.~\ref{CH3OHmixture}), which is due to a CH$_3$OH-CO ice interaction. Transmission IR spectra of a CH$_3$OH:CO mixture show a band at 2136 cm$^{-1}$ \citep{Bisschop:thesis,Palumbo:1993}. When the methanol bands grow also the band at 2135 cm$^{-1}$ increases. Figure~\ref{CH3OHmixture} shows how the peak position of CO shifts with the methanol content in the reflection spectra. The RAIR spectra on which this graph is based, are taken of ice layers that are formed by co-deposition of CO and CH$_3$OH of known ratio. The CO stretching mode in H$_2$O:CO and NH$_3$:CO mixtures shows similar behaviour \citep{Standford:1988, Bouwman:2007}. Like H$_2$O and NH$_3$, CH$_3$OH is able to form hydrogen bonds and these hydrogen bonds  most likely cause  the redshift of the CO band.
 By comparing the position of the peak in Fig.~\ref{COchange} at 2135 cm$^{-1}$ to Fig.~\ref{CH3OHmixture}, we conclude
  on the methanol fraction in the top layers assuming that the formed CH$_3$OH:CO mixture has the same spectral behaviour as the deposited mixtures. The observed data after three hours correspond to a CH$_3$OH:CO mixture of at least 90~\%. This means that the top layer of the ice is completely converted to H$_2$CO and CH$_3$OH and that no or very little additional mixing with CO occurs. For the H$_2$CO and CH$_3$OH band no spectral changes are observed during the experiments. 
  
{In order to quantify the use-up of CO and the formation of new products, we have to assume that the apparent absorption band strength is constant during an experiment, \emph{i.e.}~independent of the ice composition. \cite{Bouwman:2007} found that indeed the band strength of the 2143~cm$^{-1}$ CO feature is not affected within the experimental error by water content in H2O:CO-ice mixtures up to 4:1. The band strength is expected to behave similarly for a CO:CH$_3$OH-mixture. 
Furthermore, if the band strength would be strongly affected by the ice composition, the total ice thickness determined using a constant band strength would vary in time, whereas the real thickness is constant. Since this does not occur, we estimate that the change in band strength due to changing ice composition is negligible and well within our error bars.}

Figure~\ref{dNdt} (a) shows the time evolution of the integrated CO, H$_2$CO and CH$_3$OH signals in symbols. It shows how the amount of CO decreases as the abundance of H$_2$CO grows for four different temperatures. After bombardment with 1$\times$10$^{17}$ H atoms cm$^{-2}$ the formation of methanol kicks off at the expense of the growth of the H$_2$CO abundance. Similar abundance evolutions as a function of fluence have been reported by \cite{Watanabe:2006}. This indicates that the fluence is determined with relatively high accuracy since in both experiments different atomic sources (Tschersich vs.~microwave induced plasma) and different calibration methods are used.

\begin{figure}[h]
\includegraphics[width=0.45\textwidth]{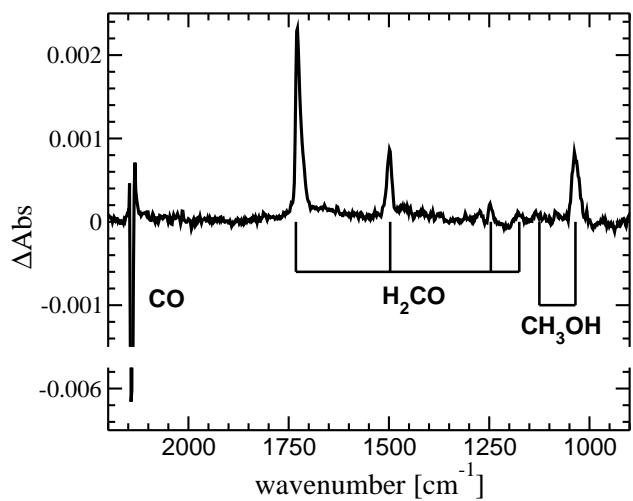}
\caption{RAIR difference spectrum of a CO ice at 12.0 K exposed to 5.4$\times$10$^{17}$ cm$^{-2}$ H atoms at a flux of 5$\times$10$^{13}$ cm$^{-2}$ s$^{-1}$. {The spectrum after CO deposition is used as the reference spectrum.} Note that the CO peak reaches an absorbance difference of -0.006.}
\label{IR_hiflux}
\end{figure}

\begin{figure}[h]
\includegraphics[width=0.45\textwidth]{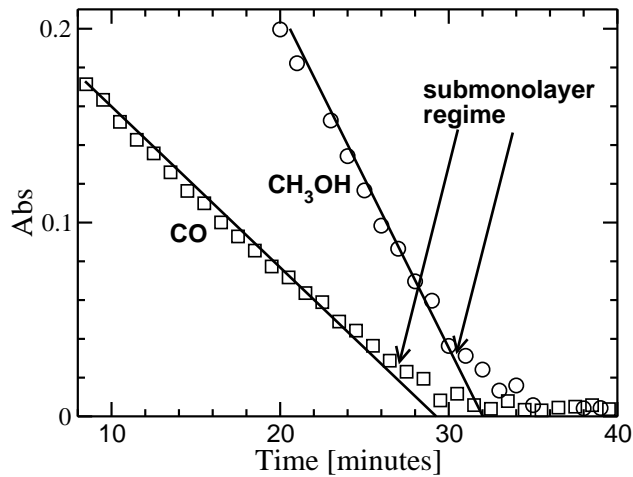}
\caption{The decrease in integrated absorbance of CO and CH$_3$OH (1035 cm$^{-1}$) following  desorption at a constant temperature of 29 and 135 K, respectively. The arrows indicate the transition points from the multi- to sub-monolayer regime.}
\label{Calibration}
\end{figure}

\begin{figure}[h]
\includegraphics[width=0.45\textwidth]{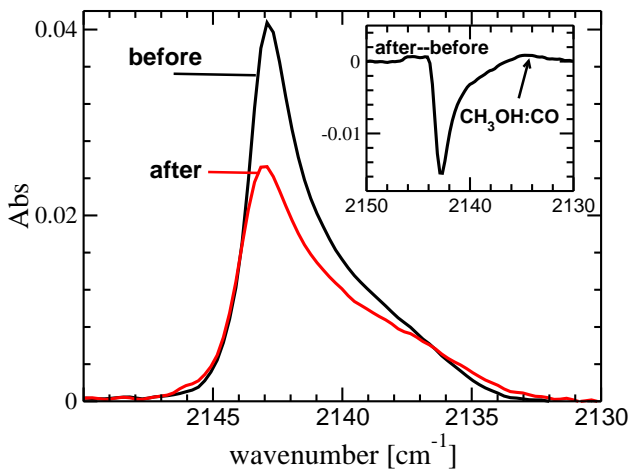}
\caption{Spectral change of the CO 2143 cm$^{-1}$ RAIR band before and after H-atom bombardment. The inset shows the corresponding difference spectrum.}
\label{COchange}
\end{figure}

\begin{figure}[h]
\includegraphics[width=0.45\textwidth]{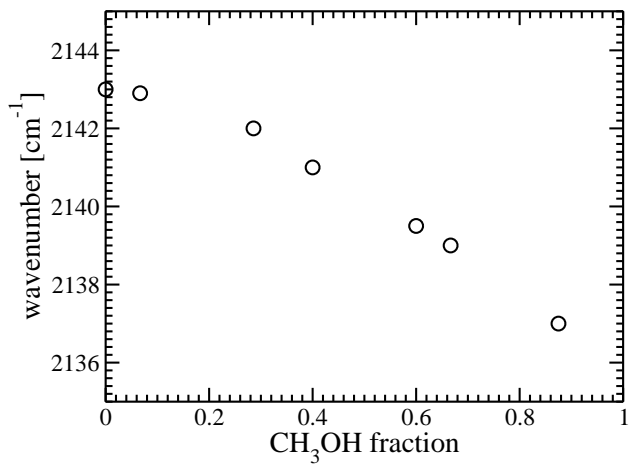}
\caption{{CO RAIR band position as a function of CH$_3$OH content in a CO:CH$_3$OH mixed ice obtained by codeposition experiments.}}
\label{CH3OHmixture}
\end{figure}

\begin{figure}[h]
\includegraphics[width=0.45\textwidth]{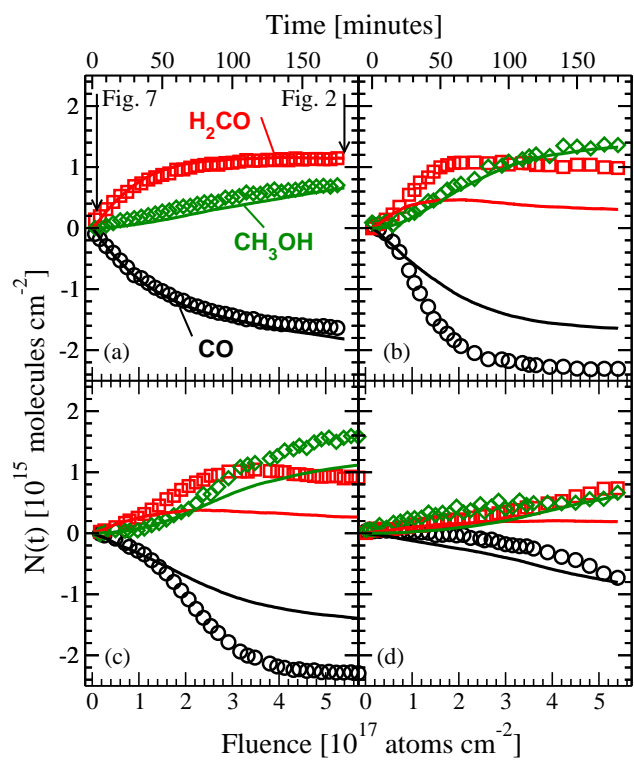}
\caption{Time evolution of the surface abundance (in molecules cm$^{-2}$) of CO, H$_2$CO and CH$_3$OH during H-atom bombardment of CO ice with a H-atom flux of 5$\times$10$^{13}$ cm$^{-2}$ s$^{-1}$ at {surface temperatures of }12.0 K (a), 13.5 K (b), 15.0 K (c), and 16.5 K (d). Experimental data (symbols) and Monte Carlo simulation results (solid lines) are shown as well. }
\label{dNdt}
\end{figure}

\subsection{Flux dependence}
As mentioned in the introduction, the apparent discrepancy between the results by \cite{Hiraoka:2002} and \cite{Watanabe:2002} was attributed to a difference in the H-atom flux used in the respective experiments. The setup in our laboratory is able to cover the entire flux range from 10$^{12}$ to 10$^{14}$ cm$^{-2}$s$^{-1}$. For high flux, both formaldehyde and methanol are formed as can be seen in Figs.~\ref{IR_hiflux} and \ref{dNdt} and in the corresponding work of \cite{Watanabe:2002}. 

\begin{figure}[h]
\includegraphics[width=0.45\textwidth]{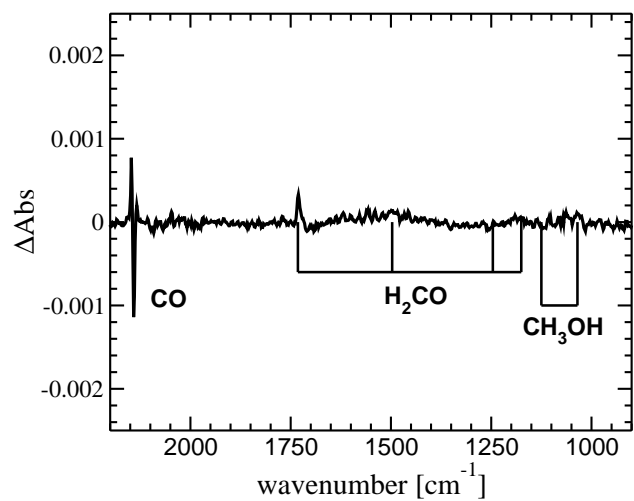}
\caption{RAIR spectrum of a CO ice at 12.0 K exposed to 1$\times$10$^{16}$ H atoms cm$^{-2}$. }
\label{IR_lowflux}
\end{figure}

\begin{figure}[h]
\includegraphics[width=0.45\textwidth]{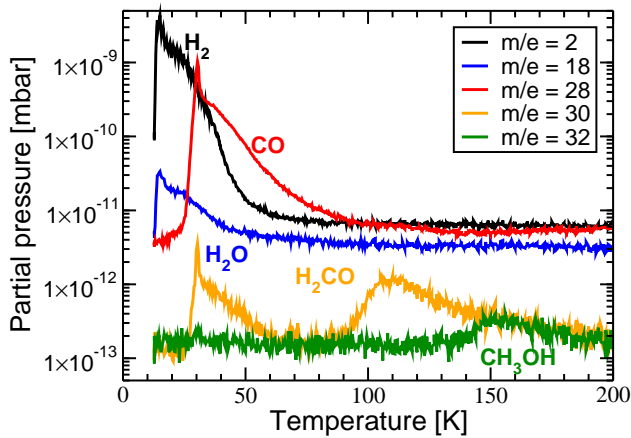}
\caption{The TPD spectra corresponding to Fig.~\ref{IR_lowflux}.}
\label{TPD_lowflux}
\end{figure}

A difference spectrum of a similar experiment but with a much lower flux of 10$^{12}$ cm$^{-2}$s$^{-1}$ is plotted in Fig.~\ref{IR_lowflux}. The exposure time here is four hours to obtain better statistics, but the total fluence of $1\times 10^{16}$ cm$^{-2}$ is still significantly less than the sample experiment shown in Fig.~\ref{IR_hiflux}. Note that the vertical scales in Figs.~\ref{IR_hiflux} and \ref{IR_lowflux} are the same. For longer exposures surface contamination will become a problem, but methanol features will eventually become detectable. As Fig.~\ref{IR_lowflux} clearly shows, much less CO is transformed to H$_2$CO and the sensitivity of the RAIR spectrometer is not high enough to confirm the formation of CH$_3$OH at these circumstances. TPD, however, is more sensitive as a diagnostics tool, although harder to use for a quantitative or time resolved analysis. Figure~\ref{TPD_lowflux}  plots several TPD spectra. It shows a small methanol desorption peak around 150 K. We have experimentally checked that the carrier of this peak is indeed formed in the ice during the hydrogen exposure and that the observed CH$_3$OH is not a contaminant in the UHV chamber. This is a strong indication that the formation mechanism of formaldehyde and methanol does not fundamentally change with varying flux. {The H$_2$O desorption at 20-30~K originates from frozen background water on the surrounding parts of the cryohead. }

Arrows in Fig.~\ref{dNdt}a indicate the corresponding fluences for the low and high flux experiments, respectively shown in Figs.~\ref{IR_lowflux} and \ref{IR_hiflux}. From this it is immediately apparent that only a limited amount of methanol can be formed under low flux circumstances. Note that \cite{Hiraoka:2002} probably used an even lower fluence since their exposure time was four times shorter than in our experiment. In addition, they used a slightly lower temperature of 10 K.

\subsection{Thickness dependence}
\begin{figure}[h]
\includegraphics[width=0.45\textwidth]{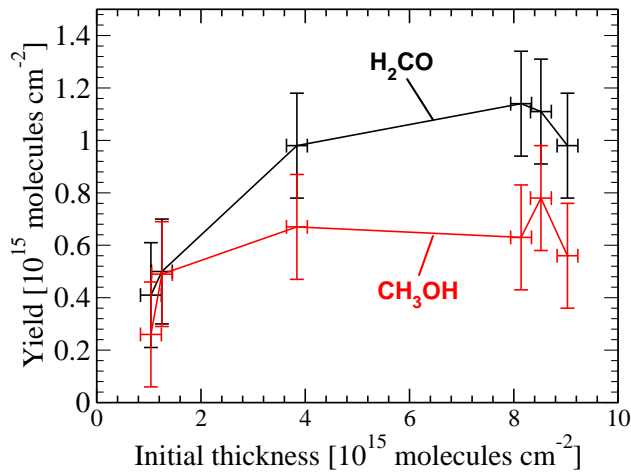}
\caption{The absolute reaction yield of H$_2$CO and CH$_3$OH after a fluence of 5.4$\times$10$^{17}$ H atoms cm$^{-2}$ as a function of the layer thickness for experiments at 12.0 K.}
\label{Yield}
\end{figure}

The effect of the initial layer thickness on the formation yield of
H$_2$CO and CH$_3$OH is investigated by repeating the sample
experiment for different  CO layer thicknesses. Figure
\ref{Yield} shows the absolute reaction yield after a fluence of
$5.4\times 10^{17}$~H~atoms~cm$^{-2}$ as a function of the layer
thickness. A steady state value for H$_2$CO is reached for this
fluence in all cases. The figure clearly shows that for CO layers
thicker than 4$\times$10$^{15}$~molecules~cm$^{-2}$ the absolute yield
is layer thickness independent and the results are reproducible within
the measurement error. The combined H$_2$CO and CH$_3$OH yield of
$2\times10^{15}$~molecules~cm$^{-2}$ is lower than the
$4\times10^{15}$~molecules~cm$^{-2}$ penetration column. From these
experiments we conclude that the penetration column of the H atoms
into the CO ice is at most $4\times10^{15}$~molecules~cm$^{-2}$ at 12.0 K. This corresponds to 4~monolayers~(ML) of solid
(bulk) CO molecules. At least half of the CO molecules in the active
layer is converted to H$_2$CO and CH$_3$OH. The
determination of the penetration column by this experiment is only an upper limit due to
the low thickness resolution in Fig.~\ref{Yield}. It is however in
agreement with the previous estimate of nearly 100~\% conversion.

\subsection{Temperature dependence}
Several experiments for different surface temperatures have been
performed. The initial layer thickness and flux values are comparable
to the values used in the sample experiment. Figures~\ref{dNdt} (b)-(d)
show the results for hydrogenation experiments at 13.5, 15.0, and
16.5 K, respectively. These clearly indicate the very different
evolution of CO, H$_2$CO, and CH$_3$OH abundance with
temperature. Table~\ref{yieldtab} gives the initial formation rate of
formaldehyde (slope at $t=0$) and the final H$_2$CO and CH$_3$OH
yields. It is also indicated whether or not a steady state is
reached. The table shows that at early times the formation rate of H$_2$CO is
much lower for higher temperatures as compared to 12.0 K.  We will
come back to this later.
The final yield of CH$_3$OH is however larger
at 13.5 and 15.0 K.  For $T>15$ K the production rate of H$_2$CO is
simply so low that a steady state is not reached. {Minimal amounts of formed methanol were also detected in experiments at 18.0 and 20.0 K, but since some CO desorption and redeposition occurs at these temperatures, they are not presented here for a quantitive discussion.}

The appearance of the extra CO band at 2135 cm$^{-1}$ indicates
that for the temperatures from 12.0 to 15.0 K a nearly pure methanol
layer is formed. We expect a similar behaviour for formaldehyde. This
means that the active CO layer involved in the reactions can be
determined directly from the steady state yield of H$_2$CO and
CH$_3$OH. This active layer increases with temperature indicating that
the penetration column of H atoms into CO ice increases with
temperature as one would expect. 
The CO molecules in the ice are more
mobile at higher temperatures making it easier for H atoms to
penetrate the CO ice, {since the ice becomes less rigid}.  Note that the absolute temperature calibration
in the set-up of Watanabe and ours appears to differ by 1-2~K
(comparing Fig.~3 in \cite{Watanabe:2006} and Fig.~\ref{dNdt} here),
but the observed trends are identical.

\begin{table*}
\caption{The production rate and yield of H$_2$CO and the yield of CH$_3$OH.}
\label{yieldtab}
\begin{center}
\begin{tabular}{cccccc}
\hline
$T$ & Rate(H$_2$CO)$_{t=0}$\tablenotemark{a}  & Yield (H$_2$CO)\tablenotemark{b} & Yield (CH$_3$OH)\tablenotemark{b} & Steady state &  Calc.~pen.~column\tablenotemark{c}\\
(K) & (10$^{-3}$ molec./H atom) & (10$^{15}$ molec.~cm$^{-2}$) & (10$^{15}$ molec.~cm$^{-2}$) & & (10$^{15}$ molec.~cm$^{-2}$)\\
\hline
12.0 & 9.0 & 1.2 & 0.8 & yes &  2.0\\
13.5 & 7.3 & 1.0 & 1.4 & yes &  2.4\\
15.0 & 3.2 & 0.9 & 1.6 & yes &  2.5\\
16.5 & 1.1 & 0.8 & 0.6 & no \\
18.0 & 1.0 & 0.5 & 0.2 & no \\
20.0 & 0.9 & 0.4 & 0.1 & no\\
\hline
\end{tabular}
\tablenotetext{a}{Rate at $t=0$ determined from slope.}
\tablenotetext{b}{Yield after three hours of H-atom exposure which corresponds to a fluence of 5.4$\times$10$^{17}$ H atoms cm$^{-2}$. The steady state yield is not reached for all temperatures (fifth column).}
\tablenotetext{c}{Penetration column obtained from columns 3, 4, and 6.}
\end{center}
\end{table*}

\section{Monte Carlo simulations}
\subsection{The method}
To infer the underlying mechanisms leading to the formation of methanol a detailed physical-chemical model is required. The present section  discusses  an approach based on the continuous time, random-walk Monte Carlo simulation. This method is different from previous studies based on rate equations and enables the study of surface processes in more detail. In addition, it gives a better understanding about what occurs physically on the surface. In contrast to an analysis using rate laws, the Monte Carlo method determines the H surface abundance by taking into account the layered structure of the ice, the H-atom flux, diffusion, reaction and desorption. This allows an extension of the results to conditions with much lower fluxes like in the interstellar medium (ISM). For a detailed description of  method and  program, see \cite{Cuppen:2007}. 

During a simulation a sequence of processes - hopping, desorption,
deposition and reaction - is performed where this sequence is chosen
by means of a random number generator in combination with the rates
for the different processes. First, an initial ice layer is created by
deposition of CO on a surface. The resulting surface roughness of this
layer depends on temperature and flux. {For the experimental conditions that are simulated here, the CO ice is compact with a maximum height difference across the surface of only 2-3 monolayers.} Hydrogen atoms and hydrogen
molecules are subsequently deposited according to their relative
abundance in the H-atom flux {with an angle perpendicular to the surface to mimic the experimental conditions.} They move, react and desorb according
to rates with a similar form as used in gas-grain models
\begin{equation}
R_x = A \exp\left(-\frac{E_x}{T}\right),
\end{equation}
where $E_x$ is the activation energy for process $X$ and $A$ is the
pre-exponential factor for which a constant number of $\nu$~$\sim$~$kT/h$~=~$2\times
10^{11}$~s$^{-1}$ is used. The activation energies are not well
determined \emph{ab initio} or by experiment.  The desorption energies are
determined by the binding energy as explained below and depend on an
energy parameter $E$.  The barriers for reaction are used as a
parameter to fit the data.  The barrier for hopping (diffusion) from
site $i$ to $j$ is obtained by 
\begin{equation}
E_{\rm hop}(i,j) = \xi E + \frac{\Delta E_{\rm bind}(i,j)}{2}.
\label{Ehop}
\end{equation}
{This expression ensures microscopic reversibility between the different types of sites.}
The parameter $\xi$ is another input parameter, which is varied between simulations. Little quantitative information is available about diffusion rates on these kind of surfaces which makes the value of $\xi$ uncertain.

Diffusion into the ice is also considered. Minimum energy path
calculations suggest that CO and H can swap position enabling an
H atom to penetrate into the CO ice (see Appendix B). The barrier for
this process strongly depends on the layer in which the H atom is
situated. In the simulations the barrier for this event is (350 +
2($z_1 + z_2$)) K for an H atom to swap between layer $z_1$ and
$z_2$. {This compares to a hopping barrier of $E_{\rm hop}^{\rm H, flat\rightarrow flat} = 256$~K and a desorption energy of $E_{\rm bind}^{\rm H, flat} = 320$~K (see next section).} \cite{Hiraoka:1998} found that hydrogen atoms can relatively
easily diffuse through the CO ice. Moreover, the current experiments
show that hydrogen atoms can penetrate into a maximum of four
monolayers for 12.0 K. Hydrogen atoms are also allowed to swap with
formaldehyde and methanol, but here the initial barrier is chosen to
be higher (450 and 500 K) since these species are heavier and are more
strongly bound in the ice matrix. 

\subsection{The CO ice layer}
Although the experimental CO layers are probably amorphous
\citep{Kouchi:1990}, crystalline layers are used in the Monte Carlo
simulations discussed here. In this way a lattice-gas Monte Carlo method can be used
which enables much longer simulation times than in off-lattice
methods. We expect the crystalline assumption to be reasonable since
the local structure of the CO layers is probably close to
crystalline. The energy that is released during deposition may help
the molecules to rearrange slightly during deposition leading to
micro-crystalline domains. The $\alpha$-CO structure
\citep{Vegard:1930} is used with layers in the (110) orientation. The
dominant faces on a CO crystal will have this crystallographic
orientation. The CO surface consists of alternating carbon and oxygen
terminated bi-layers. In the bulk configuration each CO molecule has
14 nearest neighbours:
{five in layers below, five in layers above and four in the same layer. } 
The additive energy contribution of these neighbours is
$2E$ for the layers below and $E$ for the neighbours in the same layer
or with lower $z$, the depth in layers with respect to the top
layer. The different treatment for sites below the particle is to add
a contribution for longer range interactions from the ice layer. For
atomic hydrogen, $E$ is chosen to be 32 K and for CO to be 63 K. This leads
to a binding energy of $E_{\rm bind}^{\rm H, flat} = 320$ K for H on
top of flat CO ice layer and $E_{\rm bind}^{\rm H, layer} = 448$ K and $E_{\rm bind}^{\rm CO, layer} = 882$ K for, respectively, H and CO embedded in a CO
layer. These values agree very well with binding energies obtained by
calculations with accurate H-CO and CO-CO potentials of 320, 440, and
850 K, respectively (see Appendix B).

\subsection{Comparison to the experiment}
The solid lines in Fig.~\ref{dNdt} represent the results from the
Monte Carlo calculations. The exact
mechanisms included in these simulations are discussed in more detail
in the following sections. The resulting time evolution series are in
very good agreement for 12.0 K. The agreement for 13.5, 15.0 and 16.5
K is much less. This is probably due to missing mechanisms that
promote the penetration into the ice. In the current simulations only
swapping of species is included. Because of thermal motion of the CO
molecules, ``real'' penetration in which the H atoms penetrate in the
CO matrix may be possible as well. The shape of the curves is
reproduced and only the H$_2$CO abundance levels off at too low
yields.

The main parameters varied to fit the experimental data are the
reaction barriers and the diffusion rates. The best fitting barriers
are summarized in Table~\ref{Energies}. Since the intermediate species HCO
and H$_3$CO are not experimentally detected, the barriers for
hydrogenation of these species are significantly lower than for the
other two reactions, presumably even zero. The HCO and
H$_3$CO abundances stay below detectable levels in the simulations.  The reaction
barriers for H + CO and H + H$_2$CO are temperature dependent and
increase with temperature. Our values are in good absolute agreement
with the barriers found by \cite{Awad:2005}, who also found a similar
temperature behaviour. Their values were obtained using a rate
equation analysis for $T=10$, 15 and 20~K using the data from
\cite{Watanabe:2006}. The temperature dependence suggests that there
is a clear tunnelling component for the reaction at low
temperature. The two barriers for forming H$_2$CO and CH$_3$OH show
different temperature dependencies. The formation of methanol becomes
relatively more important for higher temperature. Note that the Monte Carlo
method automatically treats a reaction in competition with desorption
and hopping. This is in contrast with gas-grain codes where it has to
be included explicitly. In order to describe the chemical processes
properly one should introduce this competition in the gas-grain model. 

\begin{table}[h]
\caption{Reaction rates and barriers for CO + H and H$_2$CO + H for different temperatures.
\label{Energies}}
\begin{center}
\begin{tabular}{ccccc}
\hline
$T$ & \multicolumn{2}{c}{CO + H} & \multicolumn{2}{c}{H$_2$CO + H}\\
& barrier & rate  & barrier & rate  \\
(K) & (K) & (s$^{-1}$) &(K) & (s$^{-1}$)\\
\hline
12.0 & 390 $\pm$ 40 & $2 \times 10^{-3}$ & 415 $\pm$ 40 & $2 \times 10^{-4}$ \\
13.5 & 435 $\pm$ 50 & $2 \times 10^{-3}$ & 435 $\pm$ 50 & $1 \times 10^{-3}$\\
15.0 & 480 $\pm$ 60 & $5 \times 10^{-3}$ & 470 $\pm$ 60 & $1 \times 10^{-3}$ \\
16.5 & 520 $\pm$ 70 & $4 \times 10^{-3}$ & 500 $\pm$ 70 & $1 \times 10^{-3}$ \\
\hline
\end{tabular}
\end{center}
\end{table}
{The errors in the energy barriers reflect the errors due to the uncertainties in the sticking probability, H-atom flux, diffusion and exact structure of the CO ice.}

{Molecular hydrogen is formed on the surface with efficiencies ranging from 3~\% ($T =16.5$~K) to 70~\% ($T =12.0$~K). However, because of the large excess energy of the formation reaction the majority of the formed H$_2$ molecules leaves the surface and the H$_2$ surface abundance is predominantly determined by impinging H$_2$ molecules.}

\subsection{Effect of diffusion}
\begin{figure}[h]
\includegraphics[width=0.45\textwidth]{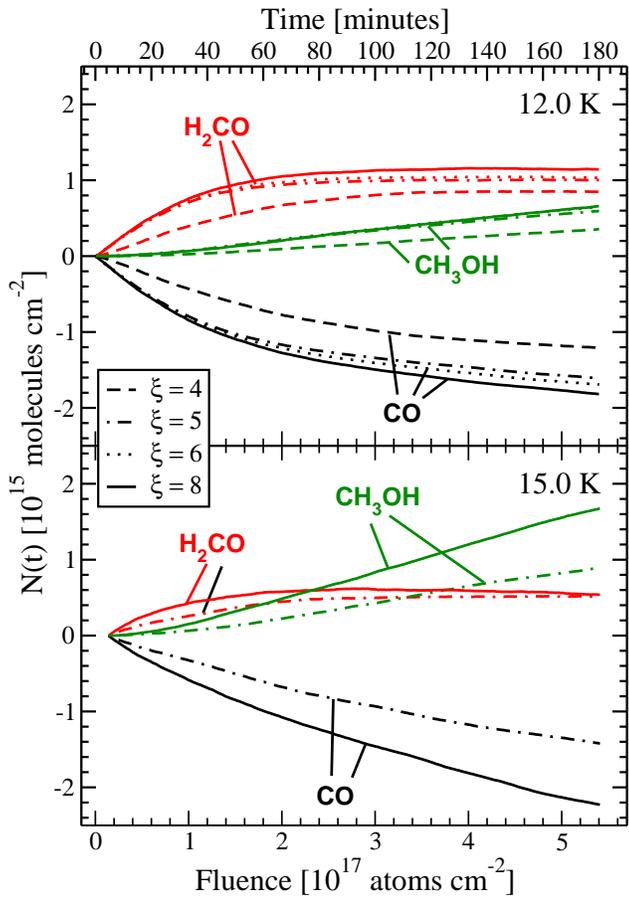}
\caption{Monte Carlo simulations of the time evolution of the surface abundance of CO, H$_2$CO and CH$_3$OH during H-atom bombardment of CO ice at 12.0 K (top) and 15.0 K (bottom). Reaction barriers for H + CO and H + H$_2$CO can be found in Table~\ref{Energies}. The diffusion is varied via the parameter $\xi$ (Eq.~5). }
\label{Xi}
\end{figure}

Since the diffusion rates are uncertain, this section discusses its
effect in more detail. Minimum energy path calculations of the
diffusion of a single hydrogen atom on a CO (110) surface (see
Appendix B) results in energy barriers ranging from 70 to 170~K ($\xi$
= 2-5.3) depending on the direction of diffusion. The Monte Carlo
program only considers one type of diffusion between ``flat''
sites. This corresponds better to the isotropic nature of an amorphous
surface. Amorphous surfaces are usually more corrugated than
crystalline surfaces which increases the hopping barrier. The second
term in Eq.~\ref{Ehop} ensures microscopic reversibility.  Figure
\ref{Xi} shows the influence of the diffusion parameter $\xi$ on the
H$_2$CO and CH$_3$OH production. The simulations are carried out in
the presence of H$_2$ for 12.0 K (top) and 15.0 K (bottom). The
difference in diffusion appears to have a larger effect for 15.0 K
than for 12.0 K. Faster diffusion (smaller $\xi$) clearly results in
less CH$_3$OH and H$_2$CO production, since the H atoms are more
likely to find each other and to react away to H$_2$. Slower
diffusion allows the H atoms more time per CO encounter to cross the
reaction barrier and to form HCO.  
In the simulations presented in Figs.~\ref{dNdt} and \ref{ISM}, we
use $\xi = 8$ to reduce the simulation time. This parameter
choice results in a ratio $E_{\rm hop}({\rm flat},{\rm flat})/E_{\rm
  bind} ({\rm flat})$ of 0.78, which is in agreement with the
experimentally found ratio for H atoms on olivine and amorphous carbon
\citep{Katz:1999}. The amorphocity of the surface may be responsible
for such a high ratio.

\subsection{Effect of H$_2$ molecules on the hydrogenation}
All simulations include deposition of both H atoms and H$_2$
molecules, which results from the undissociated H$_2$ molecules in the
H-beam.  
If the H$_2$ molecules are  excluded from the
simulations, the formation of H$_2$CO and CH$_3$OH is affected in only
a limited set of cases: fast diffusion and high temperature. The
presence of H$_2$ appears to have mainly two effects. It limits the
penetration into the ice and it slows down the H atoms, since they
move through a ``sea'' of H$_2$. The first has a negative effect on
the production rate, the latter depends on the reaction barrier.

The experimental results at temperatures higher than 12.0 K show
non-first-order behaviour at early times (exponential decay of
CO). The H$_2$CO production rate increases until 30 and 50 minutes of
exposure for $T$ = 13.5 and 15.0 K, respectively. After this time, the
H$_2$CO and CH$_3$OH follow the expected first order behaviour. None
of the simulations in Fig.~\ref{Xi} shows this trend. The only
mechanism which is able to describe this phenomenon is an increasing
effective H-atom flux with time. This increasing effective flux can be
due to an increased sticking of atomic hydrogen on the surface. Since
the incoming H atoms are relatively warm, they need to
dissipate this extra energy to the surface in order to stick. Because
CO is relatively heavy as compared to the H atoms, this energy
dissipation will be inefficient and most of the H atoms will scatter
back into the gas phase. Once the surface abundance of the much
lighter H$_2$ molecules increases, the sticking of the H atoms to the
surface will increase as well. We assume a 1~\% sticking for H atoms
and H$_2$ molecules on a bare CO surface and a 65~\% sticking of
H atoms on a surface which is fully H$_2$ covered . The sticking probability is
further assumed to grow linearly with H$_2$ coverage. 
The H$_2$ surface abundance reaches a contant level of 0.39~ML after a few minutes for $T$ = 12.0 K. This results in a sticking of H atoms to the surface of 26~\%.  For higher
temperatures it takes noticeably longer to equilibrate, explaining the
non-linear behaviour at early times and indicating a lower final sticking probability. The solid lines in
Fig.~\ref{dNdt} include this mechanism.

As mentioned earlier, \cite{Watanabe:2002} concluded that the
temperature of the beam has little effect on the hydrogenation
process, which seems to contradict our H$_2$ argument. However, their
experiments were carried out at 10 K, where the surfaces will be
covered with hydrogen atoms early on in the experiment because of the
enhanced sticking at low temperatures. They further
reported an unknown flux difference between the cold and warm beam,
which makes quantifying the sticking probability using these
experiments not possible. In conclusion, the temperature of the beam
can affect the effective flux of H atoms landing on the surfaces, but
it does not introduce additional energetic effects which influence the
crossing of the barrier.

\section{CO hydrogenation under interstellar conditions}
Based on the fitting results in the previous section, the Monte Carlo
routine can now be used to simulate CO hydrogenation reactions under
interstellar conditions.  An important ingredient is the H-atom
density in the cloud. Just as in our laboratory beam, the gas in dense
clouds consists of a mix of H and H$_2$.  Under steady-state
conditions, the balance of the rates of H$_2$ formation on grains and
H$_2$ destruction by cosmic rays gives an H-atom density around 1
cm$^{-3}$ \citep{Hollenbach:1971}. This H-atom number density is
independent of the total density because both the formation and
destruction rates scale with density. Before steady-state is reached,
however, the H-atom density may be higher because the time scale for H
to H$_2$ conversion is long ($\sim 10^7$ yr) starting from a purely
atomic low density cloud \citep{Goldsmith:2007}. Our model assumes a
constant H-atom density of 10 cm$^{-3}$.  Our other model parameters are a
gas temperature of 20 K and dust temperatures of 12.0 and 16.5 K. A CO surface is
then simulated for $2\times 10^5$ yr, which corresponds to a fluence of
$10.8\times 10^{17}$ atoms cm$^{-2}$.  Note that half of this fluence was
actually realized in our experiments.
Because the H-atom velocities are low, the sticking of H atoms to the
CO ice is kept constant at 100~\%.

{
The starting configuration for the simulations is a layer of pure CO ice. This is believed to be representative for the top layers of the grain mantles in the center of a high-density collapsing cloud. Here, the ice layer is observed to consist of predominantly CO ice as the result of ``catastrophic'' CO freeze-out \citep{Pontoppidan:2006,Pontoppidan:2008}. More heterogeneous ice layers are formed at lower densities, where CO and H$_2$O are mixed, or towards the center of proto-stellar envelops or proto-planetary disks where the dust has been heated and CO has desorbed from the top layers. }

\begin{figure}[h]
 \includegraphics[width=0.45\textwidth]{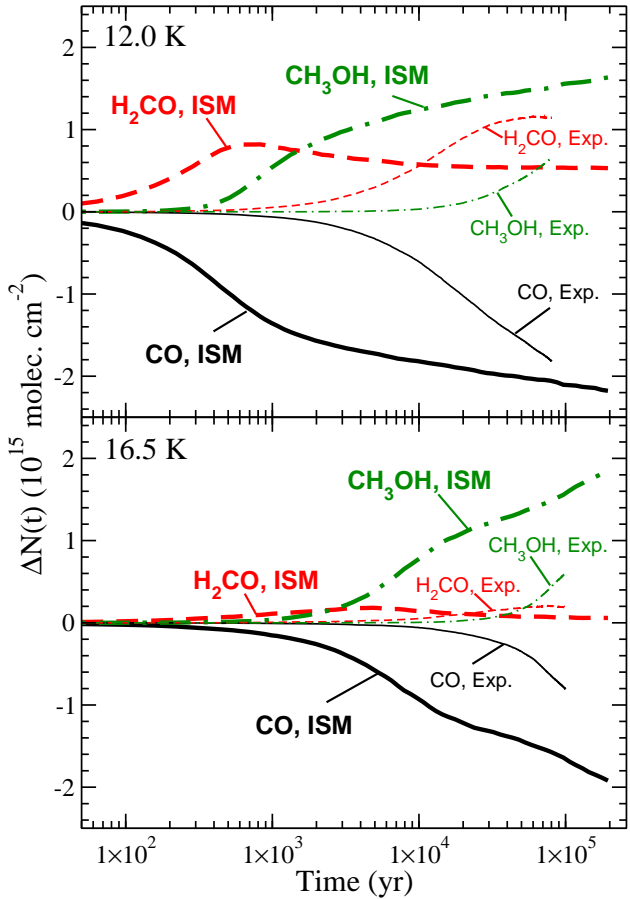}
 \caption{Monte Carlo simulations of CO-ice hydrogenation at 12.0 (top) and 16.5 K (bottom). A
   constant atomic hydrogen gas phase density of 10 cm$^{-3}$ and a
   gas temperature of 20 K is assumed. Thick lines represent
   interstellar conditions, thin lines are the scaled experimental
   simulations. The results are shown as the change in column density compared
   with t=0 yr.}
 \label{ISM}
 \end{figure}

Figure~\ref{ISM} (top) shows the resulting time evolution of CO, H$_2$CO,
and CH$_3$OH ice (thick lines) for 12.0 K.  The thin lines in Fig.~\ref{ISM}
represent the direct scaling of the simulations of the experiment to
interstellar time scales. The H$_2$CO/CH$_3$OH ratio for the low flux
simulation is very different from the scaled experimental
simulation. {The reason for this is that 
in the laboratory environment twice as many hydrogen atoms react with each other to form
H$_2$ than are involved in the four CO hydrogenation reactions since the surface density is relatively high. For interstellar
conditions the CO hydrogenation reactions are dominant and only $<5$~\% of
the reacting H atoms are converted to H$_2$. } A second effect that changes the
time evolution in the ISM is the difference in sticking. Under
laboratory conditions the sticking probability is much lower since the
incoming H atoms with room temperature cannot release their energy
very efficiently to the CO ice. The presence of H$_2$ on the surface
may have a positive effect on the sticking probability. In the ISM the
incoming atoms are much colder and energy dissipation will not be a
limiting factor for the sticking of H atoms to CO ice. This can be
modelled using the Monte Carlo simulations but only after deriving the energy
barriers by fitting the laboratory data.

\begin{figure}[h]
 \includegraphics[width=0.45\textwidth]{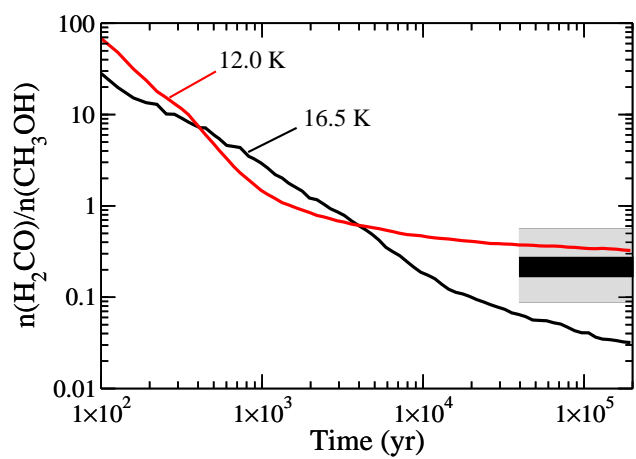}
 \caption{The H$_2$CO/CH$_3$OH ratio as a function of time obtained from the Monte Carlo simulations of CO hydrogenation at 12.0 and 16.5 K under ISM conditions (see Fig.~\ref{ISM}). The gray box indicates Spitzer ice observations, the black box gas phase observations.}
 \label{H2CO/CH3OH}
 \end{figure}

The bottom panel in Fig.~\ref{ISM} shows similar trends for 16.5 K. Again the onset of H$_2$CO and CH$_3$OH formation is at much lower fluences as compared to the experiment. At the end of the simulation nearly all H$_2$CO has been converted to CH$_3$OH. This is in contrast with the 12.0 K simulations where a constant non-zero amount of H$_2$CO remains after $2\times 10^5$ yr. The crossover point from H$_2$CO-rich to CH$_3$OH-rich ice occurs at slightly later times at 16.5 K compared to 12.0 K. This can clearly be seen in Fig.~\ref{H2CO/CH3OH} which plots the H$_2$CO/CH$_3$OH ratio for both temperatures. At early times this ratio is similar for 12.0 and 16.5 K. At $t > 10^3$~yr, the ratio starts to level off for 12.0 K, while it still decreases rapidly for 16.5 K. The noise in the curve for 16.5 K below $t = 5\times 10^3$ yr is due to the low abundances of H$_2$CO and CH$_3$OH. 

In space, the H$_2$CO/CH$_3$OH ice ratio has been determined directly for only
three high-mass young stellar objects (YSOs): W33A, NGC~7538~IRS9 and
AFGL~70009S, with inferred ratios ranging from 0.09 to 0.51
\citep{Keane:2001,Gibb:2004}.  The laboratory curves for the H$_2$CO
and CH$_3$OH production show that H$_2$CO is more or equally abundant
during most of our experiments. Thus, values as low as 0.09-0.51 cannot
easily be reproduced in the experiments.  However, the Monte Carlo
simulations for interstellar conditions have a crossover from
H$_2$CO-rich to CH$_3$OH-rich ice at significantly earlier times than
the experimental curves and a H$_2$CO/CH$_3$OH ratio of 0.51 is obtained
after $5\times 10^3$ yr at $T_{\rm dust}$=12 K. Grains at higher temperatures
will have this crossover at even earlier times and for grains with $T_{\rm dust}$=16.5 K a H$_2$CO/CH$_3$OH ratio of even 0.09 is obtained after $2\times 10^4$ yr. Thus, the observed
ratios are in agreement with the models discussed above for chemical
time scales $>2 \times 10^4$ yr, which is consistent with the estimated ages of
these high-mass protostars of a few $10^4-10^5$ yr \citep{Hoare:2007}.

CH$_3$OH ice has also been detected toward low-mass YSOs with
abundances ranging from $<$1~\% to more than 25~\% with respect to
H$_2$O ice \citep{Pontoppidan:2003,Boogert:2008}. An interesting
example is the Class 0 protostar \protect{Serpens SMM~4}, for which
a particularly high CH$_3$OH abundance of 28~\% with respect to
H$_2$O ice has been deduced for the outer envelope
\citep{Pontoppidan:2004}. The upper limit on the H$_2$CO-ice abundance
implies a H$_2$CO/CH$_3$OH ratio $ < $0.18, implying an age $> 1 \times 10^4$ yr at 16.5 K. This is consistent with the estimated time scale
for heavy freeze out in low-mass YSOs of $10^{5\pm 0.5}$ yr, including
both the pre-stellar and proto-stellar phases \citep{Jorgensen:2005I}.

Other observational constraints come from sub-millimetre observations
of the gas in a sample of massive hot cores, where a constant ratio of
H$_2$CO/CH$_3$OH of 0.22$\pm0.05$ was found
\citep{Bisschop:2007III}. If both the observed H$_2$CO and CH$_3$OH
have just evaporated freshly off the grains and if they have not been affected
by subsequent gas-phase chemistry, the observed ratio should reflect
the ice abundances. This ratio is roughly consistent with the asymptotic value that is reached in the 12 K model. This remarkably constant abundance ratio
would imply very similar physical conditions (dust temperatures,
H-atom abundances,~...) during ice formation.

In contrast, the fact that the CH$_3$OH ice abundance with respect to
that of H$_2$O is known to vary by more than an order of magnitude
suggests the opposite: that local conditions and time scales do play a role. Note, however, that for CH$_3$OH abundances as large as 25~\% (columns as large as $10^{18}$ cm$^{-2}$), the CH$_3$OH layer is approximately 25 ML thick ( $0.25 \times n({\rm H_2O}) / (n_{\rm dust} \times \textrm{$<$binding sites per grain$>$}) = 0.25 \times 10^{-4} / (10^{-12} \times 10^6) = 25$ ML), much more than can be
produced from just the upper 4 ML of the CO ice. Thus, conversion of
CO to CH$_3$OH ice must in these cases occur simultaneously with the
freeze-out and building up of the CO layer.  Pure CO ice can also
easily desorb as soon as the protostar heats up. This complicates the
use of CH$_3$OH/CO ice as an evolutionary probe.  A proper model of
interstellar CH$_3$OH ice formation should therefore include the
changing CO-ice abundances and dust temperatures in the pre- and
protostellar phases, taking into account the time scales for CH$_3$OH-ice formation compared with those of CO adsorption and desorption.
This paper provides the necessary molecular data to work on such a
model.

\section{Conclusion}
The present paper shows that the formation of methanol via successive
hydrogenation of CO and H$_2$CO is efficient under various laboratory
conditions covering $T_{\rm surf} = 12- 20 $ K, ice thicknesses between
1$\times 10^{15}$ and 1$\times 10^{16}$ molecules cm$^{-2}$ equivalent
to 1 and 10~ML bulk CO, and H-atom fluxes between $1\times 10^{12}$
and 5$\times 10^{13}$ cm$^{-2}$s$^{-1}$. {Our results show that the
discrepancy between \cite{Hiraoka:2002} and \cite{Watanabe:2002} was
indeed mainly due to the use of different H-atom fluxes and we agree with the latter that CH$_3$OH is
formed at low temperature.} \emph{On the basis of this, the surface
hydrogenation of CO can now be safely used to explain the majority of
the formed methanol in the interstellar medium}, where it serves as a
key molecule in the synthesis of more complex molecules.

Energy barriers for the H + CO and H$_2$CO + H reactions are obtained
by fitting Monte Carlo simulation results to the experimental
data. Using these barriers the methanol production is simulated for
interstellar conditions. The obtained H$_2$CO and CH$_3$OH abundances
do not scale directly with fluence due to a different relative
importance of H$_2$ production and CO hydrogenation in space compared
with the laboratory, {as can be clearly seen by comparing the thick and thin lines in Fig.~\ref{ISM}}. But the laboratory experiments are required to derive the necessary rates that serve as input for the Monte Carlo program. The obtained H$_2$CO/CH$_3$OH ratios for the
interstellar simulations are in closer agreement with observational
limits than direct translation of the experimental observations.

Monte Carlo simulations of the hydrogenation process show that
the presence of H$_2$ has three effects: it promotes the sticking of
the warm H atoms, it limits the penetration into the ice and it slows
down the diffusion of H atoms. The first effect will be negligble
under interstellar conditions since the incoming H atoms will be cold
already and the sticking probability will therefore be high regardless
of the substrate. The latter two effects will be important and are
similar to the conditions in the laboratory with also a high H$_2$
abundance.

The experiments show that the hydrogenation process is thickness
independent for layers thicker than 4 $\times 10^{15}$ cm$^{-2}$ and
that the active layer, which contains only a limited amount of CO
after steady state is reached, becomes slightly thicker with temperature. For
temperatures higher than 15.0 K, a clear drop in the production rate
of methanol is observed. This is probably due to two effects: the
desorption of H atoms becomes important and the sticking of H atoms is
reduced due to the low H$_2$ surface abundance. Both effects cause the
H surface abundance to drop substantially at those temperatures and
therefore reduce the probability of hydrogenation reactions to occur
in the laboratory. Simulations of CO hydrogenation in space show a strong temperature dependence of the H$_2$CO/CH$_3$OH ratio over several orders of magnitude. The CH$_3$OH abundance changes with time, temperature and fluence. 

\begin{acknowledgements}
  Part of this work was supported by the Netherlands Research School
  for Astronomy, NOVA, and Netherlands Organisation for Scientific
  Research (NWO) through a VENI grant. We
  thank  Stephan
  Schlemmer and Helen Fraser for their contribution during the first
  construction phase and Gijsbert Verdoes, Martijn Witlox and Ewie de
  Kuyper from the Fijn Mechanische Dienst for their support. Ayman
  Al-Halabi, Lou Allamandola, Eric Herbst, Xander Tielens, Klaus Pontoppidan, and Zainab Awad have contributed to this
  work through long and inspiring discussions. 
\end{acknowledgements}

\begin{appendix}

\section{Absolute and relative H-atom flux determination}
\subsection{Absolute flux determination}
The (accuracy of the) absolute value of the H-atom flux at the ice surface is obtained by estimating lower and upper limits in two independent ways.  We exemplify here the H-atom flux determination
for the case of our standard values with an H$_2$ pressure in
the chamber of $p_{\rm H_2}= 1\times 10^{-5}$~mbar and a filament temperature of $T = 2300$~K.

The lower limit on the absolute flux is directly available
from the experimental results presented in \cite{Ioppolo:2008}. That paper discusses the H$_2$O$_2$  and H$_2$O production from H-atom bombardment of O$_2$-ice in time using the same setup and settings. During the first hour, H$_2$O$_2$ and H$_2$O are produced with an almost constant production rate of  $6.0 \times 10^{12}$~molecules cm$^{-2}$~s$^{-1}$. Since both molecules contain two hydrogen atoms, this means that the H-atom flux should be at least twice this value. Assuming a conservative sticking probability of hydrogen
atoms at 300 K to O$_2$-ice at 12-28 K of at most 50~\%,
we determine a lower limit on the flux results in $2.4 \times 10^{13}$~cm$^{-2}$~s$^{-1}$.

The determination of the upper limit on the H-atom flux is more elaborate and involves several steps. Fig.~\ref{setup} shows that the hydrogen atoms travel from the source through the atomic-line chamber to a quartz pipe where the atoms are collisionally cooled and then through the main chamber onto the substrate. The final H-atom flux is then determined by
\begin{equation}
\phi_{\rm H} = \frac{N_{\rm H, source} k_1 k_2 p r}{A},
\end{equation}
where $N_{\rm H, source}$ is the number of hydrogen atoms leaving the source per second, $k_1$ is the coupling efficiency between the source and quartz pipe, $k_2$ is the coupling efficiency between the quartz pipe and the ice surface, $p$ accounts for the pressure drop between the two chambers, $r$ for the loss in H-atoms because of recombinations
in the quartz pipe and $A$ is the surface area that is exposed by the H-atom beam.

Our specific hydrogen source, used
in the experiments described here, has been tested prior to delivery at the Forschungszentrum in J\"ulich where the flux, the solid angle and dissociation rate have been measured for a wide range of H$_2$ pressures and filament temperatures. The set-up used for these calibration experiments is described in \cite{Tschersich:1998}. These measurements confirmed that there is little variation between individual instruments, since nearly identical rates have been obtained in the publications by \cite{Tschersich:1998} and \cite{Tschersich:2000} and later by \cite{Tschersich:2008} for different H-atom sources of the same type. From the flux and dissociation rate measured in J\"ulich, $N_{\rm H, source}$ can be obtained as well as $k_1$ using the solid angle information. In our example case $4.1 \times 10^{16}$~H-atoms~s$^{-1}$ leave the H-atom source and 44~\% of these atoms enter the quartz pipe which is located at a distance of 1.5 cm. 

The pipe has been designed such that the atoms cannot reach the substrate directly and that the number of hydrogen recombinations is kept to a minimum. This is achieved by using a short pipe with a high diameter/length ratio and choosing quartz which is known to have a low recombination efficiency. Following  \cite{Walraven:1982} a theoretical estimate of the number of recombinations in the pipe can be determined, considering the specific shape and material. This reduces the H-atom flux by another 27~\%. 
The pipe ends in close proximity of the cryogenic surface. The use of a pipe instead of a pinhole or a slit results in a focused H-atom beam for which the flux can be determined with relatively low uncertainty. From geometric considerations a minimum solid angle can be estimated. This will suffice, since our aim is to obtain an upper limit for the flux. The H-atom beam covers $A=4.9$~cm$^2$ of the substrate that is located 3 cm behind the quartz pipe. This spot falls completely on the surface and $k_2$ can readily be assumed to be unity. 

Finally, the pressure drop between the source and the main chamber can be determined in two ways: by a calculation using the conductance of the pipe and the pumping speed and by measuring the pressures in both chambers using undissociated beams. Both results are in reasonable agreement leading to $p = 3.2 \times 10^{-2}$.

Our upper limit for the flux is now
\begin{equation}
\phi_{\rm H} = \frac{4.1 \times 10^{16}\cdot 0.44 \cdot 1 \cdot 3.2 \times 10^{-2} \cdot 0.73}{4.9} = 8.6 \times 10^{13} {\rm cm^{-2} s^{-1}}.
\end{equation}
Deviations from this upper limit are expected to be due to a smaller $k_1$ value, because of misalignments between the source and the entrance of the quartz pipe, an underestimation of the solid angle of the exiting beam from the quartz pipe (lower $k_1$ and higher $A$), more recombinations in the pipe or backscattering of atoms from the quartz pipe to the chamber of the H-atom source.

The value for the flux adopted in the present paper is the resulting intermediate H-atom flux of $5 \times 10^{13}$~cm$^{-2}$~s$^{-1}$, which is within a factor of 2 of the upper and lower limits. It should be noted that this is a conservative error, since the actual lower and higher flux limits are likely to be higher and lower, respectively.

\subsection{Relative flux determination}
\begin{figure}[h]
 \includegraphics[width=0.45\textwidth]{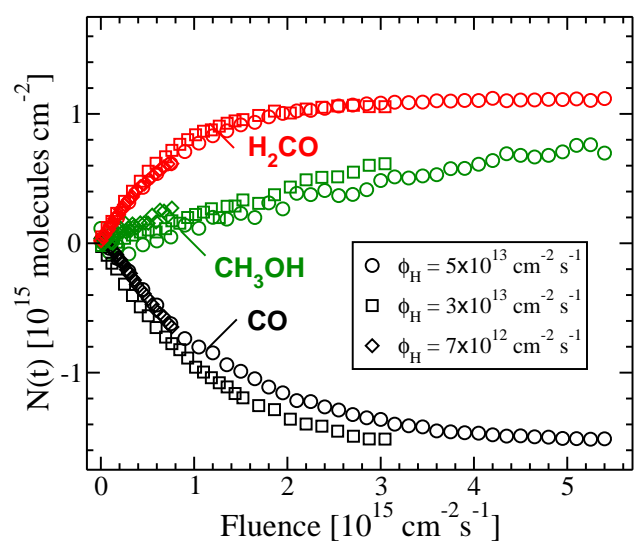}
 \caption{Time evolution of the surface abundance (in molecules cm$^{-2}$) of CO, H$_2$CO and CH$_3$OH during H-atom bombardment of CO ice at 12.0 K with three different H fluxes of 5$\times$10$^{13}$,  3$\times$10$^{13}$, and 7$\times$10$^{12}$ cm$^{-2}$. }
 \label{flux}
 \end{figure}

The accuracy in the relative flux is particularly important for the conclusion presented in this paper, more than the absolute value. For this we use the CO-hydrogenation data obtained from the experiments. Figure~\ref{flux} shows the CO, H$_2$CO and CH$_3$OH evolution as a function of fluence for three different fluxes. The fluences are calculated using the flux determination as described above. The three curves clearly overlap, which means that the accuracy of the relative fluxes is well within our error bars.
We conclude that the accuracy in the relative flux is substantially higher than the accuracy of the absolute flux, well below 50~\%. One of the main conclusions of the paper, that the discrepancy between the two Japanese groups is due to a difference in flux, as envisaged by \cite{Hidaka:2004}, is therefore solid.

 Finally, reproducing the same experiments on different days over the course of several months showed that reproducibility over periods from day-to-day to months is excellent, within a few percent.

\section{Binding energy calculations}
To calculate binding energies and barriers to diffusion, recently-developed
CO--CO and H--CO potentials are used.
Takahashi and van Hemert (in prep.)  have fitted high level electronic structure
(coupled cluster) calculations
on the CO--CO dimer to an analytic potential consisting of partial charges
on the atoms and the centres of mass of the CO molecules, atom-based
Lennard-Jones type interactions, and Morse potentials for the intramolecular
C--O
interaction. In the work by Andersson et al. (in prep.) a potential for the
interaction between a hydrogen atom
and CO has been calculated through fitting damped dispersion and exponential
repulsion potentials to coupled cluster
calculations.

Using the CO--CO potential a CO (110) surface has been created consisting of
528 CO molecules in 11 monolayers in a cell with dimensions 33.8 {\AA}
$\times$ 31.8 {\AA} in the surface plane. By applying periodic boundary
conditions an infinite surface
is created. Binding energies have been calculated by performing energy
minimisations for H atoms at different sites on top
of and inside the CO surface and comparing to the energy with the hydrogen
far away from the surface. In the same manner
the binding energy for a CO molecule in the top layer has been calculated.
In all instances the top 3 monolayers of the ice have been allowed to relax.

To calculate energy barriers to diffusion on and into the surface, initially the Nudged Elastic Band
(NEB) method \citep{Jonsson:1998} has been used to map out the
minimum energy path (MEP) connecting two potential minima. To fine-tune the
barrier height, the Lanczos
method is used
to optimize the saddle point of the potential energy \citep{Olsen:2004}.
\end{appendix}

\end{document}